\documentclass[prd, twocolumn, nofootinbib]{revtex4}

\usepackage{epsfig} 

\newcommand{\beq}{\begin{equation}}
\newcommand{\eeq}{\end{equation}}
\newcommand{\beqa}{\begin{eqnarray}}
\newcommand{\eeqa}{\end{eqnarray}}
\newcommand{\om}{\Omega_m}
\newcommand{\ome}{\Omega_e}
\newcommand{\wt}{w_{\rm tot}}

\def\la{\mathrel{\mathpalette\fun <}}

\def\fun#1#2{\lower3.6pt\vbox{\baselineskip0pt\lineskip.9pt
  \ialign{$\mathsurround=0pt#1\hfil##\hfil$\crcr#2\crcr\sim\crcr}}}

\begin{document} 

\title{Dark Energy in the Dark Ages} 
\author{Eric V.\ Linder} 
\affiliation{Berkeley Lab, University of California, Berkeley, CA 94720} 

\begin{abstract} 
Non-negligible dark energy density at high redshifts 
would indicate dark energy physics distinct from a cosmological constant 
or ``reasonable'' canonical scalar fields.  Such dark energy can be 
constrained tightly through investigation of the growth of structure, 
with limits of $\la2\%$ of total energy density at $z\gg1$ for many models. 
Intermediate dark energy can have effects distinct from its 
energy density; the dark ages acceleration can be constrained to last less 
than 5\% of a Hubble e-fold time, exacerbating the coincidence problem. 
Both the total linear growth, or equivalently $\sigma_8$, and 
the shape and evolution of the nonlinear mass power spectrum for $z<2$ 
(using the Linder-White nonlinear mapping prescription) provide important 
windows.  Probes of growth, such as weak gravitational 
lensing, can interact with supernovae and CMB distance measurements to 
scan dark energy behavior over the entire range $z=0-1100$. 

\end{abstract} 


\date{\today}

\maketitle

\section{Introduction} \label{sec:intro}

At recent times the expansion of the universe has accelerated and 
dark energy has dominated the total energy 
density.  Earlier, the universe was in a matter dominated, decelerating 
expansion epoch.  This can be seen directly through precision 
distance-redshift measurements of Type Ia supernovae (SN Ia) at 
redshifts $z\sim1$ and precision cosmic microwave background (CMB) 
measurements at $z\sim1000$, and indirectly through the presence of 
large scale structure such as galaxies and clusters of galaxies that 
require a matter dominated epoch in order to form. 

In this article we consider the extent of our knowledge about the 
behavior of dark energy in the dark ages at $z\approx 2-1000$.  This is 
sometimes phrased in terms of ``early dark energy'' -- the fraction of 
the total energy density at the CMB last scattering surface due to 
dark energy, though we will present a more general treatment.  

While canonical dark energy models with near cosmological constant 
behavior do not predict any substantial dark energy effect at $z>2$, 
limits on early dark energy are already important for cosmological parameter 
estimation.  For the parameter constraints by \cite{jarvisctio} using weak 
gravitational lensing data, \cite{linderjarvis} pointed out that a 
prior on the early dark energy fraction is needed to remove 
a second likelihood peak far from the concordance cosmology.  

We look to the growth of structure to provide a third window on the 
nature of dark energy, especially in the intermediate epoch. 
We suppose that SN Ia will provide accurate characterization of dark 
energy at $z<2$, so our focus is to probe for unexpected behavior at higher 
redshifts where canonical dark energy models have negligible influence. 
High redshift distance measures are not the appropriate probe for 
early dark energy, however, as even a drastic jump from cosmological 
constant equation of state behavior to matter behavior at $z=1.7$ 
imparts less than 1\% 
effect to the distance measured to $z=3$.  Similarly, the CMB 
temperature power spectrum does not care about the 
dark energy density as such, but rather the expansion history, so an early 
smooth dark component with an equation of state acting like matter does 
not disturb the CMB.  Therefore we concentrate on the growth of structure 
for probing dark energy in the dark ages, since growth measurements at $z<2$ 
are sensitive to changes at $z>2$, and to an early smooth dark component 
regardless of equation of state. 

In \S\ref{sec:ede} we consider early dark energy where the contribution 
to the total energy density at CMB last scattering is much larger than 
the canonical $10^{-9}$ of the cosmological constant case ($\Lambda$CDM). 
Intermediate redshift effects of ``dark age'' dark energy are treated 
in \S\ref{sec:ide}.  We discuss general limits on the dark energy 
contribution in \S\ref{sec:concl}, along with assessment of the 
measurements needed for such constraints.

\section{Early Dark Energy} \label{sec:ede}

Constraints on the amount of dark energy at early times necessarily 
depend on the dark energy model and its evolution over a wide range 
of redshifts.  This can be approached through phenomenological 
parameterizations, e.g.~along the lines of 
\cite{wetterich,dorankw,doranrobbers}. 
For their specific models they find limits of a few percent on the 
contribution of dark energy to the total energy density.  We follow 
this phemenological approach with some variations, focusing on 
the physics where possible. 

The major influence of early dark energy is on the growth of matter 
density perturbations $\delta\equiv\delta\rho/\rho$.  
(Note we consider modes 
on scales below the horizon where dark energy inhomogeneity is negligible 
due to the sound speed for quintessence being the speed of light.) 
We solve the 
linear growth equation for general dark energy evolution \cite{linjen} 
with scale factor $a=1/(1+z)$ by a 4th order Runge-Kutta scheme to find 
the growth history $g(a)=\delta/a$. 
A purely matter dominated universe has $g=1$.  Of particular interest 
are the quantities $g_0=g(a=1)$, the total growth by the present, and 
the ratio $R\equiv g(a=0.35)/g_0$.  The first measures the linear growth 
amplitude, proportional to the mass fluctuation amplitude $\sigma_8$, 
and the second provides an excellent indicator of the nonlinear power 
spectrum \cite{linwhite}, at least for canonical dark energy. 

One approach to investigating variations in growth behavior from 
early dark energy is to take 
the $z\lesssim2$ cosmology to be essentially fixed by accurate 
observations such as SN Ia distance measurements; we denote this as 
``with respect to $\Lambda$'' and keep the matter density $\om$ (and 
the present equation of state ratio $w_0$ where appropriate) 
the same for the early dark energy and the fiducial (flat, $\om=0.28$) 
$\Lambda$CDM case.  Another is to compare models where the distance 
to CMB last scattering, $d_{\rm lss}$, is held fixed, denoted ``with 
respect to CMB'' (so $\om$ is changed, and we 
can also alter the Hubble constant $h$ so as to preserve the quantity 
$\om h^2$ and hence essentially the CMB temperature power spectrum).  
We comment on the differences between the two for each model class. 

Note that the amplitude of growth today $g_0$ is degenerate 
with the present mass fluctuation amplitude $\sigma_8$, or the primordial 
density perturbation amplitude $\delta_H$.  These in turn are correlated 
with other CMB parameters such as the scalar tilt and reionization optical 
depth, so this points up the importance of CMB polarization measurements 
for fuller understanding of dark energy.

\subsection{Mocker models} \label{sec:mocker} 

The dark energy density evolution and equation of state (EOS), or 
pressure to density ratio $w(z)$ are tied 
together.  To attain a substantial level of early dark energy one 
requires the early equation of state not to be appreciably negative, 
for a monotonic evolution.  
Rather than parameterize directly in terms of the dark energy density, 
we can adopt the physical characteristics of certain dark energy equation of 
state models.  Mocker models \cite{paths} of dark energy 
have the requisite behavior, acting similar to matter ($w\approx0$) 
at early times before evolving to an accelerating component and eventually 
cosmological constant ($w=-1$) behavior.  In the equation of state phase 
plane of $w$ and $w'=dw/d\ln a$ these have the dynamics $w'=Cw(1+w)$ with 
solutions 
\beqa 
w(a)&=&-1+\left[1-\frac{w_0}{1+w_0}a^C\right]^{-1}\,,\\ 
\rho_{\rm de}(a)&=&\rho_{\rm de}(1)\,\left[(1+w_0)\,a^{-C}-w_0\right]^{3/C}. 
\eeqa 

\begin{figure}[!hbt]
\begin{center} 
\psfig{file=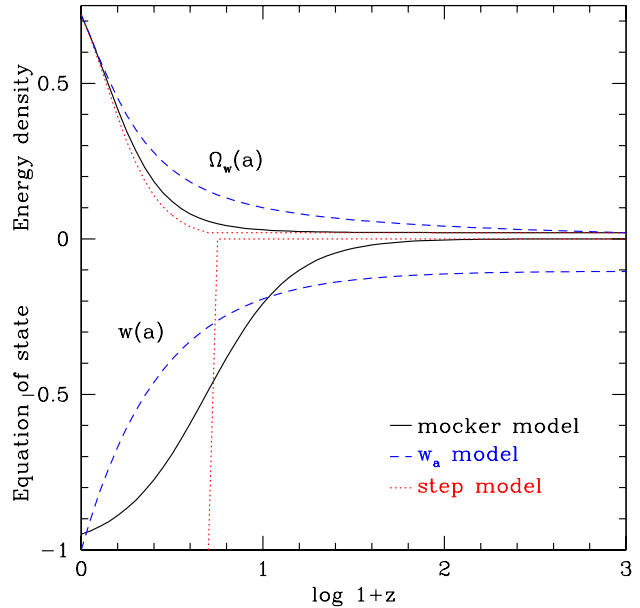,width=3.4in} 
\caption{The dark energy density, and equation of state, shown as a 
function of redshift for the three classes of models considered here. 
The fraction of dark energy density today and at the CMB last scattering 
surface ($z=1089$) are held fixed. 
}
\label{fig:dens} 
\end{center} 
\end{figure}

At early times in the matter dominated era, the mocker models have 
an energy density that scales as matter and so contribute a fixed 
fraction $\Omega_e$ of the total energy density (the dark energy 
fraction as a function of redshift is shown in Fig.~\ref{fig:dens} 
for the three classes of models considered in this section).  
Rather than using 
$\{\Omega_{\rm de},w_0,C\}$ as the parameters we can use 
$\{\Omega_m,w_0,\Omega_e\}$ and investigate the effects of $\Omega_e$, 
or the nearly equivalent $\Omega_{\rm de}(z_{\rm lss})$,  
on observational quantities.  We find that for $10^{-6}\la\Omega_e\la0.1$, 
the value of $C\approx 0.5-2$.  Interestingly, this puts the mocker model 
squarely within the freezing region of the phase space \cite{caldlin}.

\begin{figure}[!hbt]
\begin{center} 
\psfig{file=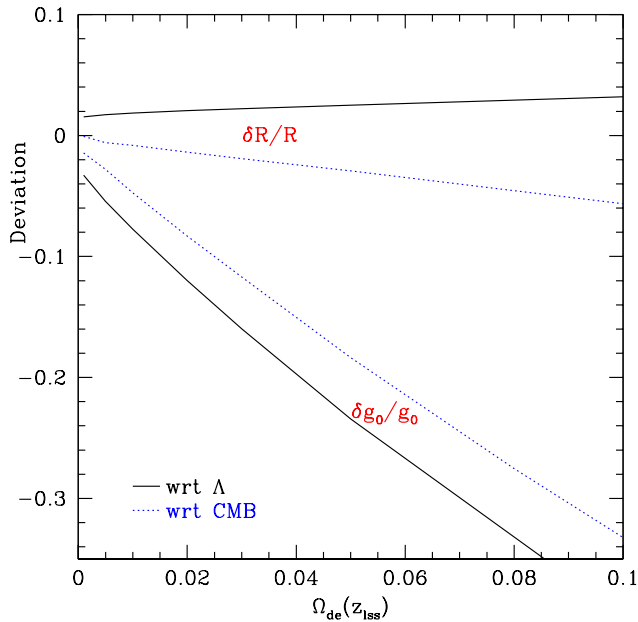,width=3.4in} 
\caption{Increasing the dark energy density at high redshifts causes 
strong deviations in the total linear growth achieved by today, $g_0$, 
as well as changes in the evolution parameter $R$, for mocker 
dark energy (here with $\om=0.28$, $w_0=-0.95$).  Solid curves show 
deviations with respect to the $\Lambda$CDM 
case with the same low redshift expansion (i.e.\ $\om$), while dotted 
curves have the same distance to last scattering as the $\Lambda$CDM case. 
}
\label{fig:gdevmock} 
\end{center} 
\end{figure}

Figure~\ref{fig:gdevmock} shows the impact on the growth measures as 
a function of the early dark energy density.  We compare the growth to 
the cosmological constant case, for the fixed low redshift 
cosmology and fixed CMB approaches.  The greatest deviation is in $g_0$, 
but this total growth can be renormalized by a shift in the mass 
fluctuation amplitude $\sigma_8$ (see \cite{doran01} for a detailed 
discussion of $\sigma_8$ in early dark energy models).  Still, values 
of $g_0$ or 
$\sigma_8$ differing from the $\Lambda$CDM value by more than 10\% are 
already disfavored \cite{seljak,abazajian,semboloni}, 
and future observations should limit the uncertainty 
to less than 3\%.  The dark energy density for a mocker model is thus 
currently limited to less than 1.5\% of the total for $w_0>-0.95$ 
(as $w_0$ approaches $-1$, the mocker model looks more and more like a 
cosmological constant for the entire matter domination era and 
constraints vanish).  If $w_0=-0.85$, then the density would be bounded 
below 0.4\% to give less than 10\% growth deviation.

Note that using the CMB matching rather than the low redshift matching 
does not significantly affect the amount of early dark energy density 
tolerated for the growth.  The low $z$ matching in turn still gives 
$d_{\rm lss}$ within 1.7\% for $\ome=0.02$ (current constraints are 
1.8\%, derived here from \cite{spergel06}).  The mocker models 
can also be distinguished relative to the constant EOS models with the 
same present value $w_0$.  
(Even for $\ome=10^{-20}$, $C=0.15$, not 0, so 
the mocker model approaches the constant EOS limit very slowly.)  

In the absence of bounds on $g_0$ or $\sigma_8$, the constraints from 
$R$ and the nonlinear power spectrum have a shallower 
dependence on $\Omega_e$.  The deviation in $R$ is at the 2-6\% level 
(for $w_0=-0.95$), or roughly a 4-12\% effect in the nonlinear power 
spectrum.  From the Linder-White \cite{linwhite} nonlinear mapping 
prescription, a variation $\Delta R/R$ around $\Lambda$CDM translates as 
\beq 
\Delta R/R\approx -0.75\,\Delta\om +0.21\,\Delta w_0+0.052\,\Delta 
w_a, \label{eq:dr} 
\eeq 
where $w(a)=w_0+w_a(1-a)$ is the standard EOS parametrization.  So a 
measured deviation of 2\% in $R$ has the equivalent effect as a change in 
the cosmological model by 0.027 in $\om$, or 0.095 in $w_0$, or 
0.38 in $w_a$.  Such a deviation again gives limits around the 2\% level in 
early dark energy density. 

While we have assumed that $z<2$ cosmology will be accurately mapped 
by SN Ia distance measurements, one should check that uncertainties 
in $\om$ and $w_0$ do not significantly blur the bounds on $\ome$. 
In fact, allowing for covariance between the parameters degrades the 
estimation of $\ome$ by less than 50\%: $\ome$ is very weakly correlated 
with the other parameters (for example the correlation coefficient 
with $\om$ is 0.12).  As discussed in \S\ref{sec:concl}, 
we make no claims for high precision in constraining $\ome$, our aim 
is factor of two, or even order of magnitude constraints, on non-canonical 
dark energy. 

\subsection{Standard models} \label{sec:wa}

\begin{figure}[!hbt]
\begin{center} 
\psfig{file=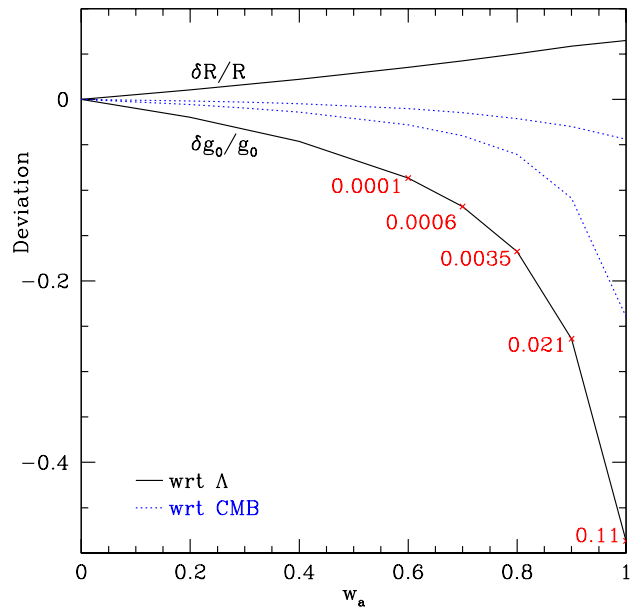,width=3.4in} 
\caption{Dark energy with EOS of the standard form $w(a)=w_0+w_a(1-a)$ 
can have a significant effect on growth even with small energy density 
at high redshift.  The dark energy not only does not contribute to 
clumping, but opposes it through a negative EOS (here $w_0=-1$).  
The influence on growth comes from both effects.  Values of 
$\Omega_{\rm de}(z_{\rm lss})$ are indicated at the red x's. 
}
\label{fig:gdevwa} 
\end{center} 
\end{figure}

One can attain early dark energy within the 
standard parametrization $w(a)=w_0+w_a(1-a)$, which has been shown 
to fit wide varieties of dark energy models, including early dark 
energy \cite{linprl}.  When the early EOS $w(a\ll1)=w_0+w_a$ 
approaches zero, the early dark energy density can contribute a 
non-negligible fraction.  Note that this is not a constant fraction 
at high redshift, unlike the mocker case, so we quote 
$\Omega_{\rm de}(z_{\rm lss})$.  Here, a 10\% deviation in $g_0$ is attained 
for a much lower $\ome$ than in the mocker case (due to the evolution 
of the density fraction, or conversely longer 
persistence of acceleration effects), except in the CMB-matched 
approach where $\ome$ is again near 1\%.  
(With the low $z$ cosmology matching instead, $d_{\rm lss}$ deviates by 
2.3\% from the $\Lambda$ case, for $w_a=0.6$.) 
Because of the increased 
dynamic range of $\ome$, we illustrate the growth deviations in $g_0$ 
and $R$ as a function of $w_a$ in Figure~\ref{fig:gdevwa}.  It is the 
presence of early dark energy density that causes the deviation, not 
any ``breakdown'' in the $w_a$ parametrization.

Recall that in Eq.~(\ref{eq:dr}) we discussed the change in the key 
growth ratio $R$ arising from a change in the cosmological parameters; 
here we see that the change $\Delta R/R$ is linear for a wide range of 
$\Delta w_a$, not just a small perturbation around $w_a=0$.  We also 
very clearly see one of the advantages of the Linder-White prescription 
\cite{linwhite}, that matching the growth matches the 
CMB temperature power spectrum.  They arranged the canonical $w_a$ 
model such that the growth ratio $R$ stayed fixed (and hence obtained 
a highly accurate nonlinear mass power spectrum) and found that the 
distance to CMB last scattering automatically was nearly perfectly 
preserved.  Here we show the converse, that matching $d_{\rm lss}$ 
almost perfectly matches the growth ratio $R$, as illustrated by the 
upper dotted line in Fig.~\ref{fig:gdevwa}.

\subsection{Step models} \label{sec:step} 

The tight bounds on early dark energy density in the previous two models 
could be a product of the slow evolution of the dark energy properties 
from matter-like to cosmological constant-like.  Here we consider a 
rapid transition, in the form of a step function where $w(z\ge z_c)=0$ 
and $w(z<z_c)=-1$.  This will preserve the property that $\ome$ is a 
constant at high redshift in the matter dominated era.  (Note that 
if one makes a step in $\Omega_{\rm de}(z)$ instead, one will have 
a singularity in $w(z)$.) 

\begin{figure}[!hbt]
\begin{center} 
\psfig{file=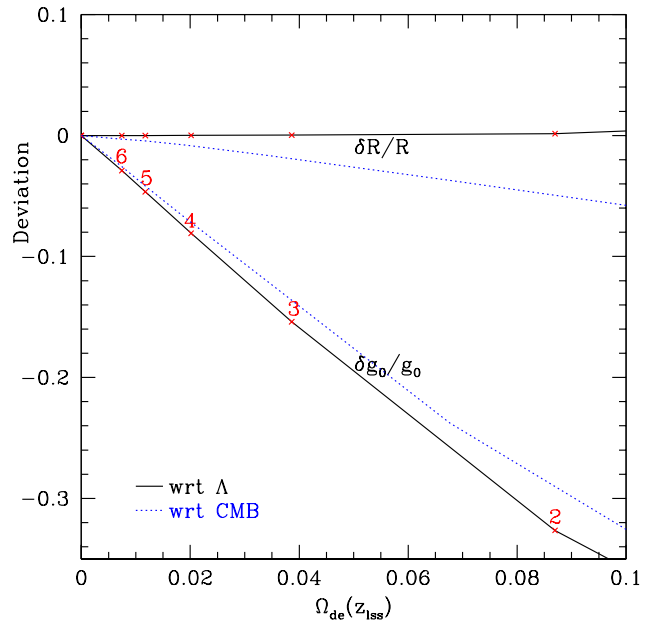,width=3.4in} 
\caption{As Fig.\ \ref{fig:gdevmock} but for stepped dark energy.  The 
EOS jumps from $-1$ at $z<z_c$ to 0 at $z\ge z_c$; this gives a 
conservative limit relative to a smoother transition or more negative 
early EOS.  The values of $z_c$ 
corresponding to $\Omega_{\rm de}(z_{\rm lss})$ are indicated at the red x's. 
}
\label{fig:gdevstep} 
\end{center} 
\end{figure}

Restricting the allowed deviation in $g_0$ to below 10\% implies that 
$\ome<2.5\%$, 
similar to the mocker models.  This also corresponds to requiring $z_c>3.65$. 
Figure~\ref{fig:gdevstep} plots the deviations in $g_0$ and $R$ with 
respect to $\Lambda$CDM, and with respect to the $\Lambda$CDM model when 
$d_{\rm lss}$ is held fixed.  Note that in the former case (matching low 
redshift cosmology) the deviations in $R$ are negligible for a transition 
$z_c>2$ since $R$ involves the ratio between growth at $z=1.86$ and 0. 
Aside from this, the constraints on $\ome$ from the two approaches are 
very similar.  The red x's are labeled with $z_c$ (purely a function of 
$\ome$), for the low redshift cosmology matching model.  

Note that even when the dark energy has $w=0$, it is not equivalent to 
matter because it is not allowed to clump.  This brings up an interesting 
point about the impact of dark energy on reducing the source term in the 
matter growth equation vs.~its impact through accelerating the expansion.  
We address this distinction in the next section.

\section{Intermediate dark energy} \label{sec:ide} 

Dark energy has two distinct effects on growth -- it changes the Hubble 
friction term by accelerating the expansion, and it changes the source 
term by reducing the fraction of the total energy density subject to 
gravitational clumping and structure formation (recall we assume it 
itself does not clump).  A more subtle consequence of these effects 
involves the indirect influence of dark energy on the growth ``velocity'' 
\beq 
f=\frac{d\ln\delta}{d\ln a}=1+\frac{d\ln g}{d\ln a}. 
\eeq 
This last effect effectively resets the boundary conditions, meaning 
an intermediate, even transient, period of acceleration can strongly 
influence later growth. 

The models of the previous section affected growth by possessing a 
substantial dark energy density at early times -- where substantial 
is relative to the $\Omega_{\rm de}(z_{\rm lss})\sim 10^{-9}$ level 
of $\Lambda$CDM.  Without early dark energy 
it is difficult to significantly affect growth through the first two 
effects.  For example, if we adapt the step model to form a box model, 
where we take $\Lambda$CDM but have a jump to $w(z)=0$ (the matter value) 
between $z_u$ and $z_d$, then we have a period where we diminish the 
source term of energy density that can clump.  This will still have an 
early dark energy density that is much smaller than one ($10^{-6}-10^{-9}$), 
and the growth is negligibly affected\footnote{An early, general analysis of 
negative equation of state and smooth components appears in \cite{linmpa88}, 
extending \cite{fry}.  The analytic formula for 
an additional smooth matter component is $g\sim a^{(-5+\sqrt{25-24\ome})/4} 
\approx a^{-3\ome/5}$.}. 
For example, with $w(z)=0$ between $z=2-20$, the deviation of $g_0$ is 
less than 10\% (recall that for the step model with $z_c=2$ the deviation 
was 33\% and $\ome=0.087$).  This sort of intermediate epoch transition 
has little observational influence. 

We {\it can\/} attain strong deviation in the growth without early 
dark energy through the velocity effect.  To investigate this we 
take a box model of dark energy, but in terms of the total equation 
of state.  This acts like $\Lambda$CDM everywhere except that between 
$z_u$ and $z_d$ the total equation of state of the universe is set to -1. 
This gives a transient epoch of strong acceleration that will shut down 
matter perturbation growth.  Once the period is over, the perturbations 
are free to grow again, but from a state where the growth velocity $f$ 
had decayed from the matter dominated value of unity.  For one indication 
of how important the growth velocity is, note that cold dark matter 
perturbations can grow in the radiation dominated epoch if given initial 
velocity: $\delta\sim \ln a$ (and indeed $\delta\sim a$ if $\wt=+1$). 

\begin{figure}[!hbt]
\begin{center} 
\psfig{file=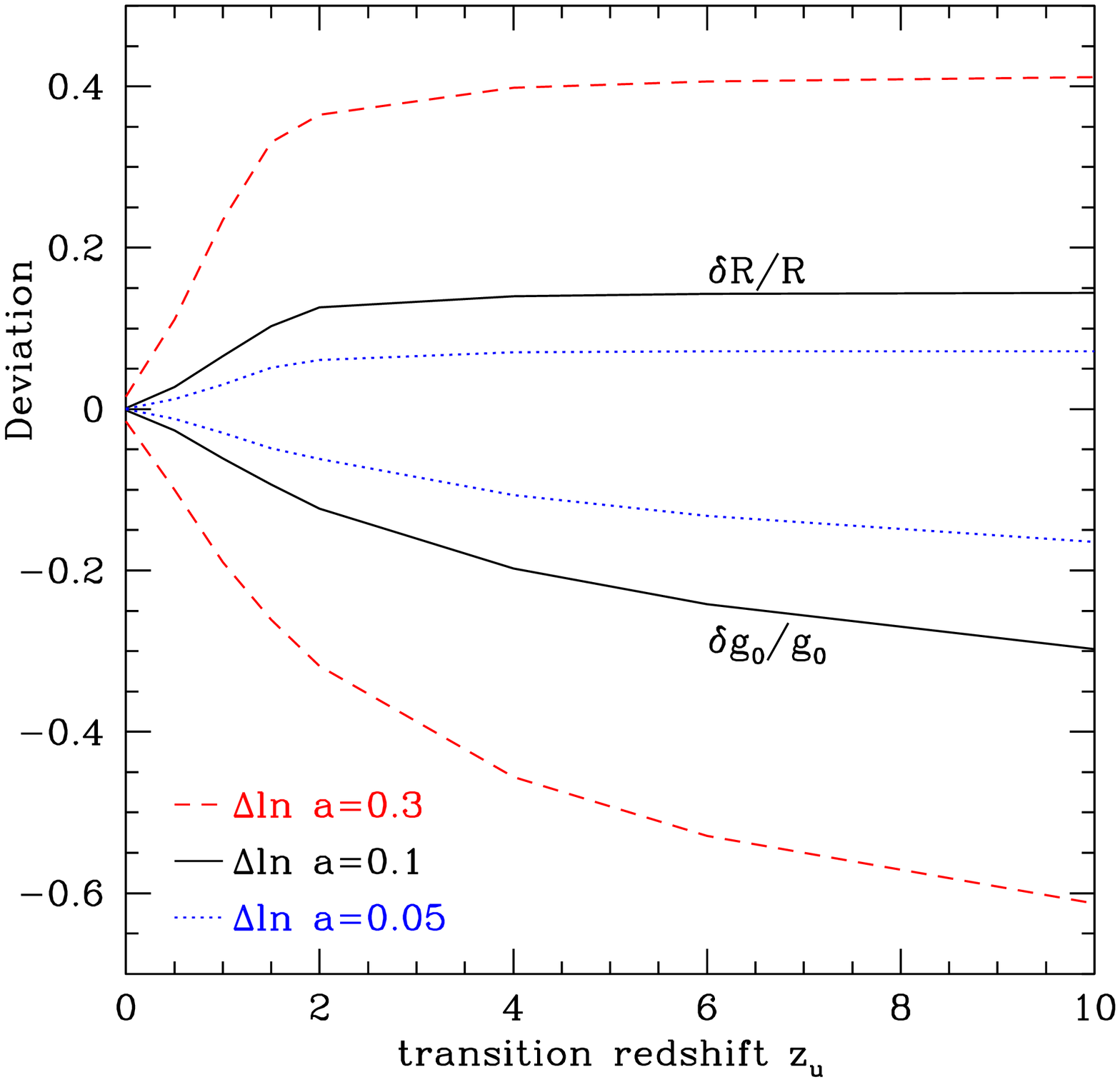,width=3.4in} 
\caption{An intermediate period of domination by dark energy in the 
dark ages can have a significant effect on growth.  This model takes 
a LCDM universe and steps the total equation of state down to $-1$ 
for a period $\Delta\ln a$ ending at $a_u$.  This halts the growth of 
structure, and even when the universe returns to matter domination 
the growth is slower than $\delta\rho/\rho\sim a$.  This shows the 
influence of the ``velocity'' of growth rather than of dark energy 
density.  
}
\label{fig:gdevbox} 
\end{center} 
\end{figure}

Figure \ref{fig:gdevbox} shows that the length of this transient 
acceleration epoch can be quite short, a fraction of an e-fold $\Delta\ln 
a$ ($\equiv\ln[(1+z_d)/(1+z_u)]$), and still cause appreciable deviation 
in growth behavior.  For a 
transition at $z_u>4$, a duration of $\Delta\ln a>0.05$, i.e.\ 5\% of that 
epoch's Hubble time, changes the present growth by more than 10\%. 
This tightly constrains models of oscillating or stochastic dark energy 
\cite{linosc,dodelstoch}, which have intermediate epochs of dark age dark 
energy 
domination\footnote{Such models attempt to ameliorate the coincidence 
problem (that dark energy is dominating during our one e-fold epoch of 
observations out of the perhaps $10^{54}$ times expansion since the 
beginning of inflation) by saying that acceleration happens periodically.}.  
Even recent 
transitions, $z_u>0.5$, would still show 10\% deviations for what 
might be considered short durations $\Delta\ln a=0.3\ll1$. 
Constraints from $R$ can be even tighter and give direct limits 
on the length of dark energy domination $\Delta\ln a$.

\section{Conclusion} \label{sec:concl} 

The nature of dark energy is so little known that we should test its 
behavior however possible, even in the ``dark ages'' $z\approx2-1000$ where 
standard models predict no effect.  The growth of large scale structure 
provides key windows on this epoch and has the potential to see early 
dark energy density or transitions in its equation of state.  Because 
of the uncertainty in the mass fluctuation amplitude, i.e.\ $\sigma_8$, 
we should employ not only the total growth factor but the growth rate, 
through the evolution of the mass power spectrum. 

For measurements of 10\% precision in total growth or, say, 5\% in the 
power spectrum evolution -- the latter related to the growth ratio 
$R=g(a=0.35)/g(1)$ through the Linder-White nonlinear mapping 
prescription -- the analysis for a variety of physical behaviors 
indicates that early dark energy density of less than $\sim2$\% of the 
total energy density does not affect observations.  This is a fairly 
conservative limit, as we have considered the extreme cases of dark 
energy behavior, e.g.\ both slow and instantaneous transitions, and 
parameter degeneracies will also degrade limits (though we find this 
is to be a surprisingly weak effect). 

Early dark energy can also affect the details of nonlinear structure, 
such as halo formation, concentration, cluster abundances, and lensing 
statistics -- see \cite{bartelmann} and references therein.  The cosmic 
microwave background can be sensitive to early dark energy (if it does 
not act like matter), but not necessarily dark age dark energy (at $z<1000$), 
since the CMB relies on the integrated dark energy density, 
which is generally dominated by low redshifts (an exception is the 
early time Sachs-Wolfe effect).  The CMB plays an important 
complementary role through constraining $\sigma_8$ for use of the total 
growth factor.  Apart from this, we find that fixing the distance to 
CMB last scattering does not greatly affect growth deviations (except 
for standard dark energy where this matching was pointed out by 
\cite{linwhite}).  

Intermediate dark energy, where there is insignificant early dark 
energy density but a dark ages period of acceleration, can be tightly 
constrained.  This impacts models that try to solve the coincidence 
problem through periodic episodes of dark energy domination.  Such 
periods must last a seemingly unnatural 5\% or less of a Hubble (e-fold) 
time.  

These analyses of physical behaviors of dark energy in the dark 
ages indicate that one can quantitatively use the observations of 
the growth of structure, 
in conjunction with accurate supernovae distance and CMB measurements, 
to have confidence that the usual scenario of dark energy as a late 
time phenomenon is valid.

\section*{Acknowledgments} 

This work has been supported in part by the Director, Office of Science,
Department of Energy under grant DE-AC02-05CH11231.

\end{document}